\documentstyle[12pt]{article}
\newcommand{\doublespace}{\renewcommand{\baselinestretch}{1.75}
\Large\normalsize}

\begin{document}
\doublespace
\begin{titlepage}

\centerline{\bf Localization of energy for a Kerr black hole}
\bigskip
\centerline{\it Jos\'e W. Maluf$\,^{*}$ and Andreas Kneip}
\centerline{\it International Centre of Condensed Matter Physics}
\centerline{\it Universidade de Bras\'ilia}
\centerline{\it C.P. 04667}
\centerline{\it 70.919-970  Bras\'ilia, DF}
\centerline{\it Brazil}
\date{}
\begin{abstract}
In the teleparallel equivalent of general relativity the energy density
of asymptoticaly flat gravitational fields can be naturally defined
as a scalar density restricted to a three dimensional spacelike
hypersurface $\Sigma$. The scalar density has a simple expression
in terms of the trace of the torsion tensor. Integration over the
whole $\Sigma$ yields the standard ADM energy. Here we obtain the
formal expression of the localized energy for a Kerr black hole.
The expression  of the energy inside a surface of constant radius
can be explicitly calculated in the limit of small $a$, the
specific angular momentum. Such expression turns out to be exactly
the same as the one obtained by means of the method proposed
recently by Brown and York.
\end{abstract}
\thispagestyle{empty}
\vfill
\par
\noindent (*) e-mail: wadih@guarany.cpd.unb.br
\end{titlepage}
\newpage

\noindent {\bf I. Introduction}\par
\bigskip
Although there is a strong belief that Einstein's
equations describe the dynamics of the gravitational field, it
has not been possible so far to arrive at a definite expression
for the gravitational energy in the context of Einstein's general
relativity. On the one hand there are prejudices based on the
principle of equivalence, which is often invoked to assure that
the gravitational energy cannot be localized\cite{Landau,MTW}.
This principle, however, is not widely accepted as an essential
element of the mathematical structure of vacuum general
relativity\cite{Synge}.
On the other hand, attempts based on the
Hilbert-Einstein action integral fail to yield an unambiguous
expression for the gravitational energy. The latter is normally
associated with surface terms in the action or in the Hamiltonian,
and these surface terms do not exhibit the appropriate transformation
properties under coordinate transformations (see the clear discussion
in \cite{Faddeev} in the case of asymptoticaly flat gravitational
fields), a fact which prevents the construction of {\it localized}
energy density.

Recently an expression for quasi-local energy has been
proposed by Brown and York\cite{Brown}. Such expression is derived
directly from the action functional $A_{cl}$. The latter is identified
as Hamilton's principal function and, in similarity  with the classical
Hamilton-Jacobi equation, which expresses the energy of a classical
solution as minus the time rate of the change of the action, the
quasilocal gravitational energy is identified as minus the proper
time rate of change of the Hilbert-Einstein action (with surface terms
included). Expressions for the quasilocal energy have been obtained
for the Schwarzschild solution\cite{Brown} and for the Kerr
solution\cite{Martinez}.

Einstein's equations can also be obtained from the teleparallel
equivalent of general relativity (TEGR). The latter ammounts to a
formulation which is geometrically different from the standard one
based on the Hilbert-Einstein action integral, but whose dynamical
content is the same. In the TEGR the metrical quantity is the
tetrad field and the action principle does {\it not}
require the addition
of non-covariant surface terms in the case of asymptotically flat
gravitational  fields. Because of this property, a {\it localized}
energy density $\varepsilon(x)$
can be naturally defined from the action principle
of the TEGR as minus the variation of the action with respect to
the proper time $N(x)$. Integration of $\varepsilon(x)$ over the
whole three-dimensional space yields the ADM energy. Moreover,
$\varepsilon(x)$ also appears in the expression of the Hamiltonian
constraint $C$ of the TEGR, a fact which allows the integral form
of $C=0$ to be written as $H-E_{ADM}=0$\cite{Maluf1,Maluf2}.

We have calculated the energy inside a sphere of radius $r_o$ in a
Schwarzschild spacetime by means of $\varepsilon(x)$\cite{Maluf1}.
The expression turns out to be exactly the same as the one obtained
through the procedure of ref.\cite{Brown} (expression (6.14) of
\cite{Brown}). In this paper we consider the Kerr black-hole.
We obtain the formal expression for the total energy in terms of
non-trivial integrals in the angular variable $\theta$. In the
limit of slow rotation (small specific angular momentum) the energy
contained within a surface of constant radius $r_o$
can be calculated. Again the result obtained here is exactly
the same as that obtained by Martinez\cite{Martinez} who adopted
Brown and York's procedure. The advantage of our procedure rests on
the fact that the localized energy associated with a Kerr
spacetime can be calculated in the general case, without recourse
to particular limits, at least by means of numerical integration,
whereas in Brown and York's procedure one has to calculate the
subtraction term $\varepsilon^0$ and for this purpose
it is necessary to
embed an arbitrary two dimensional boundary surface of the Kerr
space $\Sigma$ in the appropriate reference space ($E^3$, say),
which is not always possible\cite{Martinez}.

In section II we present the mathematical preliminaries of the TEGR,
its Hamiltonian formulation and the expression of the energy
for an arbitrary asymptoticaly flat
spacetime. In section III we carry out the
construction of triads for a three dimensional spacelike
hypersurface of the Kerr type, obtain the general expression of
the energy contained in a volume $V$ of space and provide the
exact expression of the latter  in the limit of slow rotation.
Comments and conclusions are presented on section IV.

Notation: spacetime indices $\mu, \nu, ...$ and local Lorentz indices
$a, b, ...$ run from 0 to 3. In the 3+1 decomposition latin indices
from the middle of the alphabet indicate space indices according to
$\mu=0,i,\;\;a=(0),(i)$. The tetrad field $e^a\,_\mu$ and
the spin connection $\omega_{\mu ab}$ yield the usual definitions
of the torsion and curvature tensors:  $R^a\,_{b \mu \nu}=
\partial_\mu \omega_\nu\,^a\,_b +
\omega_\mu\,^a\,_c\omega_\nu\,^c\,_b\,-\,...$,
$T^a\,_{\mu \nu}=\partial_\mu e^a\,_\nu+
\omega_\mu\,^a\,_b\,e^b\,_\nu\,-\,...$. The flat spacetime metric
is fixed by $\eta_{(0)(0)}=-1$. \\

\bigskip
\bigskip
\noindent {\bf II. The TEGR in Hamiltonian form}\par
\bigskip
In the TEGR the tetrad field $e^a\,_\mu$ and the spin connection
$\omega_{\mu ab}$ are completely independent field variables. The
latter is enforced to satisfy the condition of zero curvature.
The Lagrangian density in empty spacetime is given by

$$L(e,\omega,\lambda)\;=\;-ke({1\over 4}T^{abc}T_{abc}\,+\,
{1\over 2}T^{abc}T_{bac}\,-\,T^aT_a)\;+\;
e\lambda^{ab\mu\nu}R_{ab\mu\nu}(\omega)\;.\eqno(1)$$

\noindent where $k={1\over {16\pi G}}$, $G$ is the gravitational
constant; $e\,=\,det(e^a\,_\mu)$, $\lambda^{ab\mu\nu}$ are
Lagrange multipliers and $T_a$ is the trace of the torsion tensor
defined by $T_a=T^b\,_{ba}$.

The equivalence of the TEGR with Einstein's general relativity is
based on the identity

$$eR(e,\omega)\;=\;eR(e)\,+\,
e({1\over 4}T^{abc}T_{abc}\,+\,T^{abc}T_{acb}\,-\,T^aT_a)\,-\,
2\partial_\mu(eT^{\mu})\;,\eqno(2)$$

\noindent which is obtained by just substituting the arbitrary
spin connection $\omega_{\mu ab}\,=\,^o\omega_{\mu ab}(e)\,+\,
K_{\mu ab}$ in the scalar curvature tensor $R(e,\omega)$ in the
left hand side; $^o\omega_{\mu ab}(e)$ is the Levi-Civita
connection and $K_{\mu ab}\,=\,
{1\over 2}e_a\,^\lambda e_b\,^\nu(T_{\lambda \mu \nu}+
T_{\nu \lambda \mu}-T_{\mu \nu \lambda})$ is the contorsion tensor.
The vanishing of $R^a\,_{b\mu\nu}(\omega)$, which is one of the
field equations derived from (1), implies the equivalence of
the scalar curvature $R(e)$, constructed out of $e^a\,_\mu$ only,
and the quadratic combination of the torsion tensor. It also
ensures that the field equation arising from the variation of
$L$ with respect to $e^a\,_\mu$ is strictly equivalent to
Einstein's equations in tetrad form (we refer the reader to
refs.\cite{Maluf1,Maluf2} for additional details).

It is important to note that for
asymptoticaly flat spacetimes the total divergence in (2) does
{\it not} contribute to the action integral.  Therefore the latter
does not require additional surface terms, as it is invariant
under coordinate transformations that preserve the asymptotic
structure of the field quantities\cite{Faddeev}. In what follows
we will be interested in asymptoticaly flat spacetimes.

The Hamiltonian formulation of the TEGR can be successfully
implemented if we fix the gauge $\omega_{0ab}=0$ from the
outset, since in this case the constraints (to be
shown below) constitute a {\it first class} set\cite{Maluf2}.
The condition $\omega_{0ab}=0$ is achieved by breaking the local
Lorentz symmetry of (1). We still make use of the residual time
independent gauge symmetry to fix the usual time gauge condition
$e_{(k)}\,^0\,=\,e_{(0)i}\,=\,0$. Because of $\omega_{0ab}=0$,
$H$ does not depend on $P^{kab}$, the momentum canonically
conjugated to $\omega_{kab}$. Therefore arbitrary variations of
$L=p\dot q -H$ with respect to $P^{kab}$ yields
$\dot \omega_{kab}=0$. Thus in view of $\omega_{0ab}=0$,
$\omega_{kab}$ drops out from our considerations. The above
gauge fixing can be understood as the fixation of a {\it global}
reference frame.

Under the above gauge fixing the canonical action integral obtained
from (1) becomes

$$A_{TL}\;=\;\int d^4x\lbrace \Pi^{(j)k}\dot e_{(j)k}\,-\,H\rbrace\;,
\eqno(3)$$

$$H\;=\;NC\,+\,N^iC_i\,+\,\Sigma_{mn}\Pi^{mn}\,+\,
{1\over {8\pi G}}\partial_k (NeT^k)\,+\,
\partial_k (\Pi^{jk}N_j)\;.\eqno(4)$$

\noindent $N$ and $N^i$ are the lapse and shift functions, and
$\Sigma_{mn}=-\Sigma_{nm}$ are Lagrange multipliers. The constraints
are defined by

$$ C\;=\;\partial_j(2keT^j)\,-\,ke\Sigma^{kij}T_{kij}\,-\,
{1\over {4ke}}(\Pi^{ij}\Pi_{ji}-{1\over 2}\Pi^2)\;,\eqno(5)$$

$$C_k\;=\;-e_{(j)k}\partial_i\Pi^{(j)i}\,-\,
\Pi^{(j)i}T_{(j)ik}\;,\eqno(6)$$

\noindent with $e=det(e_{(j)k})$ and $T^i\,=\,g^{ik}e^{(j)l}T_{(j)lk}$.
We remark that (3) and (4) are invariant under global SO(3) and
general coordinate transformations.

We assume the asymptotic behaviour $e_{(j)k}\approx \eta_{jk}+
{1\over 2}h_{jk}({1\over r})$ for $r \rightarrow \infty$. In view
of the relation

$${1\over {8\pi G}}\int d^3x\partial_j(eT^j)\;=\;
{1\over {16\pi G}}\int_S dS_k(\partial_ih_{ik}-\partial_kh_{ii})
\; \equiv \; E_{ADM}\;\eqno(7)$$

\noindent where the surface integral is evaluated for
$r \rightarrow \infty$, we note that the integral form of
the Hamiltonian constraint $C=0$ may be rewritten as

$$\int d^3x\biggl\{ ke\Sigma^{kij}T_{kij}+
{1\over {4ke}}(\Pi^{ij}\Pi_{ji}-{1\over 2}\Pi^2)\biggr\}
\;=\;E_{ADM}\;.\eqno(8)$$

\noindent The integration is over the whole three dimensional
space. Given that $\partial_j(eT^j)$ is a scalar  density,
from (7) and (8) we define the gravitational
energy density enclosed by a volume V of the space as

$$E_g\;=\;{1\over {8\pi G}}\int_V d^3x\partial_j(eT^j)\;.\eqno(9)$$

\noindent In similarity with Brown and York's procedure, we can
also define the energy density as minus the variation of the action
$A_{TL}$ with respect to the proper time $N(x)$. Thus for a given
set of solutions of the classical equations of motion the energy
density $\varepsilon(x)$ can be defined as

$$\varepsilon(x)\;=\;-{{\delta A_{TL}} \over {\delta N(x)}}\;=\;
{1\over {8\pi G}}\partial_j(eT^j)\;,\eqno(10)$$

\noindent in agreement with (9). \\

\bigskip
\bigskip

\noindent {\bf III. Energy of the Kerr geometry}\par
\bigskip
The Kerr solution\cite{Kerr} describes the field of a
rotating black hole. In terms of Boyer and Lindquist
coordinates\cite{Boyer}  $(t,r,\theta,\phi)$ it is described
by the metric

$$ds^2\;=\;-{\Delta \over {\rho^2}}\lbrack dt-
a\,sin^2\theta d\phi\rbrack^2\;+\;
{{sin^2\theta}\over {\rho^2}}\lbrack(r^2 + a^2)d\phi-a\,dt\rbrack^2
\;+\;{{\rho^2}\over \Delta}dr^2\;+\;\rho^2d\theta^2\;,\eqno(11)$$

$$\Delta\; \equiv \;r^2-2mr+a^2\;,$$

$$\rho^2 \; \equiv \; r^2+a^2\,cos^2\theta\;;$$

\noindent $a$ is the specifc angular momentum defined by
$a\,=\,{J\over m}$. The components of the metric restricted to
the three dimensional spacelike hypersurface are given by
$g_{11}={{\rho^2}\over \Delta}$, $g_{22}=\rho^2$ and
$g_{33}={{\Sigma^2}\over {\rho^2}}sin^2\theta$, where $\Sigma$ is
defined by

$$\Sigma^2\;=\;(r^2+a^2)^2\,-\,\Delta\,a^2\,sin^2\theta\;.$$

We define the triads $e_{(k)i}$ as

$$e_{(k)i}\;=\;
\pmatrix{  {\rho \over \sqrt{\Delta}}sin\theta\,cos\phi &
\rho cos\theta\,cos\phi &
-{\Sigma \over \rho}sin\theta\,sin\phi \cr
{\rho \over \sqrt{\Delta}}sin\theta\,sin\phi &
\rho cos\theta\,sin\phi &
{\Sigma \over \rho}sin\theta\,cos\phi \cr
{\rho \over \sqrt{\Delta}}cos\phi &
-\rho sin\theta& 0 \cr } \eqno(12)$$

\noindent $(k)$ is the line index and $i$ is the column index. The
one form $e^{(k)}$ is defined by

$$e^{(k)}\;=\;e^{(k)}\,_rdr\,+\,e^{(k)}\,_\theta d\theta\,+\,
e^{(k)}\,_\phi d\phi\;,$$

\noindent from what follows

$$e^{(k)}e_{(k)}\;=\;{\rho^2 \over \Delta}dr^2\,+\,
\rho^2 d\theta^2\,+\,{\Sigma^2 \over \rho^2}sin^2\theta d\phi^2$$

\noindent We also obtain $e=det(e_{(k)i})=
{{\rho \Sigma }\over \sqrt{\Delta}}sin\theta$. Therefore the triads
given by (12) describe the components of the Kerr solution
restricted to the three dimensional spacelike hypersurface.

One readily notices that there is another set of triads that yields
the Kerr solution, namely, the set which is diagonal and whose
entries are given by the square roots of $g_{ii}$. This set is
not appropriate for our purposes, and the reason can be
understood even in the simple clase of flat spacetime.
In the limit when both $a$ and $m$ go to zero (12) describes
flat space: the curvature tensor {\it and} the torsion tensor
vanish in this case. However, for the diagonal set of triads
(again requiring $a\rightarrow 0$ and $m \rightarrow 0$),

$$e^{(r)}=dr\;,\;e^{(\theta)}=r\,d\theta\;,\;
e^{(\phi)}=r\,sin\theta\, d\phi\;,$$

\noindent some components of the torsion tensor do not vanish,
$T_{(2)12}=1$, $T_{(3)13}=sin\theta$, and $E_g$ calculated out
of the diagonal set above diverges when integrated over
the whole space. Therefore the use of (12) is mandatory in the
present context.

The components of the torsion tensor can be calculated in a
straightforward way from (12). Only $T_{(3)13}$ and $T_{(3)23}$
are vanishing. The others are given by:

$$T_{(1)12}\;=\;cos\theta cos\phi\,({r \over \rho}+
{a^2 \over {\rho \sqrt{\Delta}}}\,sin^2\theta -
{\rho \over {\sqrt{\Delta}}})$$

$$T_{(1)13}\;=\;sin\theta sin\phi
\lbrace -{1\over {\rho \Sigma}}[2r(r^2+a^2)-a^2 sin^2\theta\,(r-m)]+
{{r\Sigma}\over \rho^3}+{\rho \over {\sqrt{\Delta}}}\rbrace$$

$$T_{(1)23}\;=\;cos\theta sin\phi \, \lbrace
\rho-{\Sigma \over \rho}+
a^2 sin^2\theta\,({\Delta \over {\rho \Sigma}}-{\Sigma \over \rho^3})
\rbrace$$

$$T_{(2)12}\;=\;cos\theta sin\phi\,({r \over \rho}+
{a^2 \over {\rho\sqrt{\Delta}}}\,sin^2\theta-
{\rho \over \sqrt{\Delta}})$$

$$T_{(2)13}\;=\;-sin\theta cos\phi\, \lbrace
-{1\over {\rho \Sigma}}[2r(r^2+a^2)-
a^2 sin^2\theta\,(r-m)]+
{{r\Sigma} \over \rho^3}+{\rho \over \sqrt{\Delta}} \rbrace$$

$$T_{(2)23}\;=\;-cos\theta cos\phi\,\lbrace \rho -
{\Sigma \over \rho}+a^2\sin^2\theta\,({\Delta \over {\rho \Sigma}}-
{\Sigma\over \rho^3}) \rbrace$$

$$T_{(3)12}\;=\;sin\theta\,[-{r \over \rho}+
{\rho \over \sqrt{\Delta}}+
{a^2 \over {\rho\sqrt{\Delta}}}\,cos^2\theta]$$

In order to evaluate (9) we need to obtain $T^i$.
After a long calculation we arrive at

$$T^1\;=\;{\sqrt{\Delta} \over {\rho^2}}\;+\;
{\sqrt{\Delta} \over \Sigma}\;-\;
{\Delta \over {\rho^2 \Sigma^2}}\lbrack 2r(r^2+a^2)\,-\,
a^2sin^2\theta(r-m)\rbrack \;,$$

$$T^2\;=\;sin\theta\,cos\theta\,{a^2 \over {\rho^4}}\,+\,
{1\over{\rho\Sigma}}\,{{cos\theta}\over{\sin\theta}}\,\lbrack\rho
-{\Sigma\over \rho}+a^2sin^2\theta({\Delta\over{\rho\Sigma}}-
{\Sigma\over{\rho^3}})\rbrack\;,$$

$$T^3\;=\;0\;.$$

The gravitational energy density inside a volume $V$ of a three
dimensional spacelike hypersurface of the Kerr solution can now
be easily calculated. It is given by

$$E_g\;=\;{1\over {8\pi}}\int_V dr\,d\theta\,d\phi
\biggl\{ {\partial \over{\partial r}}\biggl[ sin\theta \lbrack \rho+
{\Sigma \over \rho}-{\sqrt{\Delta} \over{\rho\Sigma}}\,
\biggl(  \;2r(r^2+
a^2)-a^2sin^2\theta(r-m)\; \biggr)\rbrack \biggr]$$

$$+{\partial\over{\partial \theta}}\biggl[
{{\Sigma a^2}\over{\sqrt{\Delta}\rho^3}}\,sin^2\theta cos\theta+
{cos\theta \over \sqrt{\Delta}}
\biggl(\;\rho-{\Sigma \over \rho}+a^2sin^2\theta(
{\Delta\over{\rho\Sigma}}-{\Sigma\over{\rho^3}}\,)\;\;
\biggr)\biggr]  \biggr\} \eqno(13) $$

Next we specialize $E_g$ to the case when the volume $V$ is
contained within a constant radius surface $r=r_o$, assuming
$r_o$ to be greater that the outer horizon
$r_+\,=\,m+\sqrt{m^2-a^2}$. The
integrations in $\phi$ and $r$ are trivial. Also, because
we integrate $\theta$ between 0 and $\pi$, the second line
of the expression above vanishes. We then obtain

$$E_g\;=\;{1\over 4}\int_0^\pi d\theta\,sin\theta \biggl\{
\rho+{\Sigma \over \rho} -
{\sqrt{\Delta} \over {\rho \Sigma}}\,\biggl( 2r(r^2+a^2)-
a^2sin^2\theta
(r-m)\, \biggr) \biggr\}_{r=r_o}\;.\eqno(14)$$

We have not managed to evaluate exactly the integral above.
However, in the limit of slow rotation, namely, when
${a\over r_o}\,<<\,1$ all integrals have a simple
structure and we can obtain the
approximate expression of $E_g$. It reads

$$E_g\;=\;r_o\biggl( 1-\sqrt{1-{{2m} \over r_o}+
{a^2\over r_o^2}}\biggr)
\;+\; {a^2\over {6r_o}}\,\biggl[ 2+{{2m} \over r_o}+
\biggl( 1+{{2m}\over r_o}\biggr)
\sqrt{1-{{2m} \over r_o}+{a^2\over r_o^2}}
\biggr]\eqno(15)$$

\noindent This is exactly the expression found by
Martinez\cite{Martinez} for the energy inside the surface of
constant radius $r_o$ in a spacelike hypersurface of a Kerr
black hole, in the limit of small specific angular momentum.
As in ref.\cite{Martinez}, we have not expanded the square root
which appears in (15) in powers in $a^2 \over r_o^2$.

Let us mention finally that the expansion of $\rho+
{\Sigma \over \rho}$ in the integrand of (14)
yields $-\varepsilon_0$,
whereas the remaining term corresponds exactly to
$\varepsilon$, expressions (3.17) and (3.1) respectively
of \cite{Martinez}.  It does not seem to be possible,
however, to split $\partial_i(eT^i)$ into two terms such that
their integrals arise in the form $\varepsilon -
\varepsilon_0$.  \\

\bigskip
\bigskip
\noindent {\bf Comments}\par
\bigskip

The gravitational energy $E_g$
defined by (14) can be evaluated for an arbitrary value
of $a$ by means of numerical integration. This is the major
advantage of our procedure as compared to that of
Brown and York\cite{Brown}. By means of the latter one
cannot construct expressions like (13) and (14), which may
be useful in the study of astrophysical problems, since
in a general situation  Brown and York's procedure requires the
embedding of an arbitrary two dimensional boundary surface
of the Kerr space in the reference space $E^3$,
a construction which is not possible in general\cite{Martinez}
(the evaluation of $\varepsilon_0$ in \cite{Martinez} is only
possible in the limit ${a\over r_o}\,<<\,1$). Therefore the
present approach is more general than that of ref.\cite{Martinez}.
Finally we remark that we expect expressions (9) and (10) to be
useful in the study of the thermodynamics of self-gravitating
systems, where the gravitational energy plays the role of the
thermodynamical internal energy that is conjugate to the
inverse temperature. We hope to come to this issue in the
future.   \\

\bigskip
\bigskip
\noindent {\it Acknowledgements}\par
\noindent This work was supported in part by CNPQ.

\end{document}